\documentclass[12pt]{article}

 \usepackage{amssymb}
\usepackage{amsfonts}
 \usepackage{amsmath}
 \usepackage{graphicx}

\newcommand{\be}{\begin{equation}}
\newcommand{\ee}{\end{equation}}
\newcommand{\bea}{\begin{eqnarray}}
\newcommand{\eea}{\end{eqnarray}}

\newcommand{\p}[1]{(\ref{#1})}

\newcounter{rown}

\begin{document}

\begin{center}

{\LARGE\bf Supersymmetric Calogero models }

\vspace{0.2cm}

{\LARGE\bf from superfield gauging}

 \vspace{0.7cm}

{\large\bf E. Ivanov$^{1}$,\,\, O. Lechtenfeld$^{2}$,\,\,  S. Fedoruk$^1$}

\vskip 0.5cm

\ $^1${\it  N.N. Bogoliubov Laboratory of Theoretical Physics, \\ Joint Institute for Nuclear Research, \\
141980 Dubna, Moscow region, Russia}, \\
{\tt eivanov@theor.jinr.ru, fedoruk@theor.jinr.ru}

\vskip 0.1cm

\ $^3${\it Institut f\"{u}r Theoretische Physik and Riemann Center for Geometry and Physics, \\
Leibniz Universit\"{a}t Hannover  \\
30167 Hannover, Germany}, \\
{\tt olaf.lechtenfeld@itp.uni-hannover.de}

\end{center}

\vspace{0.2cm}

\centerline{\bf Abstract}
\begin{center}
{\parbox{15cm}{\small Using the superfield gauging procedure, we
construct new ${\cal N}\,{=}\,2$ and ${\cal N}\,{=}\,4$ superfield
systems that generalize Calogero models. In the bosonic limit,
these systems yield rational Calogero models and hyperbolic
Calogero-Sutherland models in the ${\cal N}\,{=}\,2$ case, and their
${\rm U}(2)$ spin generalization in the ${\cal N}\,{=}\,4$ case.}}
\end{center}

\setcounter{footnote}{0}
\setcounter{equation}{0}

\section{Introduction}

Calogero models \cite{Calogero69,Moser,Poly06} are text-book examples of integrable multi-particle one-dimensional ($d\,{=}\,1$) systems.
The simplest is the so-called rational Calogero model
\begin{equation}\label{Calog-N0}
S_{C} = \frac12 \int dt \Big[ \sum_{a} \dot x_a \dot x_a - \sum_{a\neq b} \frac{c^2}{4(x_a - x_b)^2}\,\Big]\,, \qquad a,b=1,\ldots,n\,,
\end{equation}
which describes the interaction of $n$ identical particles with a potential inversely proportional to the square of the distance and invariant with respect to transformations of the $d\,{=}\,1$ conformal group $\mathrm{SO}(1,2)$
\begin{equation}\label{Calog-tr}
\delta t = \alpha\,, \qquad \delta x_a = \frac12 \dot\alpha x_a\,, \qquad \partial_t^3\alpha = 0\,.
\end{equation}
The Calogero-Moser system \cite{Calogero69,Moser,Poly06} is a generalization of the system (\ref{Calog-N0}) by adding an oscillator term
$\sim  \sum_{a\neq b} (x_a - x_b)^2$.
Being interesting, in the first turn, as integrable systems, rational Calogero models also bear relationships with superstring theory and M-theory \cite{GibT99,IKN04}.

Besides conformally invariant systems, some other many-particle integrable Calogero-type models are known \cite{OP1}, e.g., Calogero-Sutherland hyperbolic systems \cite{Calogero69,Moser,Suther71,Poly06}
\begin{equation}\label{CalogS-N0}
S_{CS} = \frac12 \int dt \Big[ \sum_{a} \dot q_a \dot q_a - \sum_{a\neq b} \frac{c^2}{4 \sinh^2 \frac{q_a - q_b}{2}}\,\Big],
\end{equation}
and their trigonometric analogues.

A natural generalization of the Calogero and Calogero-Sutherland systems is their supersymmetric variants.
An ${\cal N}\,{=}\,2$ superextension was built in \cite{FrM}, where each bosonic coordinate $x_a$
was completed, by two fermionic fields, to the multiplet $({\bf 1,2,1})$. Thus, the model contains $n$ physical bosons and $2n$ fermions.
The corresponding ${\cal N}\,{=}\,2$, $d\,{=}\,1$  superfield action in the limit of zero fermions is reduced to the action of the rational Calogero model.
Similarly, one can set up  ${\cal N}\,{=}\,2$ extension of the Calogero-Sutherland models \cite{DLM}.
Passing to supersymmetric extensions with higher ${\cal N}$ encounters certain problems.
E.g., when generalizing to the ${\cal N}\,{=}\,4$ case, the coordinate set $\{x_a\}$ must be enlarged to a set of $({\bf 1,4,3})$ supermultiplets with $n$ bosonic and $4n$
fermionic physical fields  \cite{Wyl}.
However, when constructing the corresponding superfield action,
which should yield the potential of the $n$-particle Calogero system in the bosonic sector,
there arise two prepotentials connected by a set of nonlinear differential equations \cite{BGL05} including the
WDVV equations \cite{W90,BVV91}, explicit solutions to which are known only for small values of~$n$.

There is another type of supersymmetrization, in which the above problems do not arise,
although the models constructed in this way are ``non-minimal'': they contain ${\cal N}n^2$ fermions for each set of $n$ bosonic coordinates \cite{FIL09,FIL12,FIL19}.
This supersymmetrization is based on the gauging method \cite{DI06},
developed previously in \cite{Poly91,GN94,Poly06} in application to the Calogero bosonic systems.
A particular Calogero model arises as a result of eliminating gauge fields in the Lagrangian of some matrix gauge-invariant system.
In this talk, based on the results of \cite{FIL09,FIL12,FIL19}, we expound how this approach can be applied
to some ${\cal N}\,{=}\,2$ and ${\cal N}\,{=}\,4$ superfield matrix models in order to obtain
new versions of supersymmetric Calogero models.

\section{Calogero and Calogero-Sutherland models as gauge models}

To illustrate the method we use, let us first show how one can reproduce the well-known conformal mechanics model \cite{AFF},
\begin{equation}\label{AFF-act}
S_0= \int dt \,L_0\,,\qquad L_0=\frac12 \left(\dot x^2 -c^2\, x^{-2}\right)\,,
\end{equation}
from a different $d\,{=}\,1$ system with gauge symmetry \cite{FIL12}.
Consider the model of complex field $v(t)$ with the Lagrangian
\begin{equation}\label{La-z}
L_v = \frac12 \, \dot{v}\,\dot{\bar v} +\frac{im}{2}\, \left(\dot v \bar v -  v\dot{\bar v}\right) \,,
\end{equation}
which is invariant under global transformations $v' = e^{-i\lambda} v$, $\bar{v}' = e^{i\lambda} \bar{v}$.
Now we extend the Lagrangian \p{La-z} so that it possesses gauge symmetry with a local parameter: $\lambda \rightarrow \lambda(t)$.
To accomplish this, we introduce $d\,{=}\,1$ gauge field $A(t)$ that lengthens the derivatives as
$\dot{v} \rightarrow \nabla v=\dot{v} + iA v$, $\dot{\bar v} \rightarrow \nabla \bar v=\dot{\bar v} - iA \bar v$.
The resulting system with the Lagrangian
\begin{equation}\label{gauge-z}
L^g_v = \frac{1}{2}\, \nabla v \nabla \bar v + \frac{im}{2}\,  \left(\nabla v \bar v -  v\nabla \bar v\right) + c\, A
\end{equation}
is invariant, up to a total derivative, under the gauge transformations introduced above,
supplemented by the transformation $A' = A + \dot\lambda$.
The last term in \p{gauge-z}, with the constant $c$, is also gauge-invariant up to a total derivative.
It is an analogue of the well-known Fayet-Iliopoulos term.

Choosing the gauge $v\,{=}\,\bar v\,{\equiv}\,x(t)$ and eliminating the field $A(t)$ by its equation of motion,
we obtain the following expression for the Lagrangian in this gauge
\begin{equation}\label{L-g-fix}
L_{gauge}= \frac{1}{2}\,\dot x^2 - \frac{1}{2}\left(m x - c x^{-1}\right)^2 \, .
\end{equation}
For $m\,{=}\,0$ it coincides with the Lagrangian \p{AFF-act}.
We note that the initial action with the Lagrangian \p{La-z} for $m\,{=}\,0$, as well as the gauge-invariant model \p{gauge-z},
are invariant under the conformal transformations $\mathrm{SO}(1,2)$ \p{Calog-tr}, supplemented by the transformations $\delta A(t)\,{=}\, - \dot{f}\, A(t)\,$.
As a result, the action with the Lagrangian \p{AFF-act} also has $d=1$ conformal symmetry.

In the gauge approach, the Calogero system is described by $\mathrm{U}(n)$-invariant matrix system \cite{Poly91,GN94,Poly06},
incorporating the $n\times n $ Hermitian matrix field $X_a^b$, $a,b=1,\ldots ,n$; the complex $\textrm{U}(n)$-spinor field $Z_a(t)$, $\bar Z^a = (Z_a)^*$
and $n^2$ Hermitian gauge fields $A_a^b$. The gauge-invariant action has the form
\begin{equation}\label{fed}
S_C =
\frac12\int dt  \,\Big[\, {\rm tr}\left(\nabla\! X \nabla\! X \right) + i\, (\bar Z \nabla\!
Z -
\nabla\! \bar Z Z) + 2c\,{\rm tr} A  \,\Big]\, ,
\end{equation}
where the following definitions are used for the covariant derivatives
\begin{equation}\label{cov-der}
\nabla\! X = \dot X +i [A, X]\,, \qquad \nabla\! Z = \dot Z + iAZ\,, \qquad \nabla\!
\bar Z
= \dot{\bar Z} -i\bar Z A \,.
\end{equation}
The action \p{fed} is invariant under local $\mathrm{U}(n)$ transformations acting on the spinor indices $a,b$ of all involved quantities,
with the matrix field $A_a^b$ as a gauge field.
Using $n^2 - n$ local transformations, we can fix the gauge $X_a{}^b\,{=}\,0$ with $a\neq b$.
The residual gauge transformations generated by the abelian subgroup $[\mathrm{U}(1)]^n$ are then fixed
by the reality conditions $\bar Z^a\,{=}\,Z_a$, $Z_a$ being subject to the constraints $Z_a Z_a\,{=}\,c$ for each $a$.
As a result, after eliminating the auxiliary and gauge fields, the action \p{fed} is reduced to the action \p{Calog-N0} of the Calogero model.
Since the original action \p{fed} is conformally invariant, the model \p{Calog-N0} is also conformally invariant.

The Calogero-Sutherland model can be deduced by a similar gauging procedure from the system involving
a nonlinear kinetic term of the sigma-model type for the matrix field $X_a^b$:
\begin{equation}\label{L-CS-N0}
S_{CS} =
\frac12\int dt  \,\Big[\, {\rm tr}\left(X^{-1}\nabla\! X X^{-1}\nabla\! X \right) + i\, (\bar Z \nabla\!
Z - \nabla\! \bar Z Z) + 2c\,{\rm tr} A  \,\Big] \,.
\end{equation}
Following the same pattern as in the rational case, we arrive at the action
\begin{equation}\label{L-CS-N0a}
S_{CS} = \frac{1}{2}\,\int dt  \,\Big[\, \sum_{a} \frac{ \dot x_a \dot x_a}{(x_a)^2}  - \sum_{a\neq b} \frac{x_a x_b c^2}{(x_a -
x_b)^2}\,\Big] \,,
\end{equation}
which, in terms of the variables $q_a\,{=}\,\ln x_a $, coincides with \p{CalogS-N0}.
Like the initial matrix action, the resulting action does not possess conformal invariance.

\section{${\cal N}\,{=}\,2$ Calogero and Calogero-Sutherland models}

To construct ${\cal N}\,{=}\,2$ supersymmetric generalization, we will resort to the same strategy,
proceeding now from the matrix ${\cal N}\,{=}\,2$ superfields and effecting a superfield gauging procedure.
The input superfield set involves $n\times n$ matrix Hermitian superfield with components ${\mathcal X}_a{}^b(t, \theta,\bar\theta)\,, \;a,b=1,\ldots ,n$,
describing $n^2$ supermultiplets $({\bf 1, 2, 1})$, and a chiral $\mathrm{U}(n)$-spinor superfield
${\mathcal Z}_a (t_{L}, \theta)$, $\bar{\mathcal Z}^a(t_R, \bar\theta)$,
$\bar D {\mathcal Z}_a=0$, $D \bar {\mathcal Z}^a=0$,
$t_{\!\scriptscriptstyle{L,R}}=t\mp i\theta\bar\theta$.
The free action for these superfields,
\begin{equation}\label{L-C-N2}
S^{N=2} = \frac{1}{2}\int dt d\theta d\bar\theta \,\big[\, {\rm tr} \left( \bar D {\mathcal X}\,  D {\mathcal X}\, \right) - \bar{\mathcal Z}\,
{\mathcal Z}\big],
\end{equation}
remains invariant under global $\mathrm{U}(n)$-transformations
${\mathcal Z}^{\prime}\,{=}\,  e^{i\lambda} {\mathcal Z}$,
$\bar {\mathcal Z}^{\prime}\,{=}\,  \bar{\mathcal Z}\, e^{-i\bar\lambda}$,
${\mathcal X}^{\,\prime}\,{=}\, e^{i\lambda}\, {\mathcal X}\, e^{-i\bar\lambda}$.
Gauging these symmetries amounts to passing to the chiral and antichiral superfield parameters $\lambda$ and $\bar\lambda$.
To ensure invariance, the Hermitian gauge superfield $V$ is introduced, with the transformation law:
$e^{2V^{\,\prime}}\,{=}\, e^{i\bar\lambda}\, e^{2V} \, e^{-i\lambda}$.
The gauge-invariant action has the form
\begin{equation}\label{S-C-N2}
{S}^{N=2}_{C} = \frac{1}{2}\int dt d^2\theta \,\Bigg[\, {\rm tr} \left( \bar{\mathcal D}
{\mathcal X}
\, e^{2V} {\mathcal D} {\mathcal X}\, e^{2V} \right) - \bar {\mathcal Z}\, e^{2V}\!
{\mathcal Z} +2c\,{\rm tr} V \,\Bigg]\,,
\end{equation}
where covariant derivatives are defined as
\begin{equation}\label{cov-der-N2}
{\mathcal D} {\mathcal X} =  D {\mathcal X} + e^{-2V} (D e^{2V}) \, {\mathcal X} \,, \qquad
\bar{\mathcal D} {\mathcal X} = \bar D {\mathcal X} - {\mathcal X} \, e^{2V} (\bar D
e^{-2V})\,.
\end{equation}
It can be shown that the initial matrix action \p{L-C-N2} and its gauge-invariant analogue \p{S-C-N2} possess ${\cal N}\,=\,2$ superconformal symmetry $\mathrm{SU}(1,1|1)$.

Using the component expansions
${\mathcal X} = X + \ldots$,
${\mathcal Z} = Z + \ldots$,
choosing the Wess-Zumino gauge $V = \theta\bar\theta A(t)$ and eliminating auxiliary fields, we obtain the following component action
\begin{equation}\label{S-C-N2c}
S^{N=2}_{C} =  \frac{1}{2}{\displaystyle\int} dt \,\Big[{\rm
tr}\,\nabla X \,\nabla X  + i\,(\bar Z \nabla Z - \nabla \bar Z Z) +2c\,{\rm tr} A  \, +
\,i\,{\rm tr} (\bar\Psi \nabla \Psi - \nabla \bar\Psi \Psi)\Big].
\end{equation}
Here $\nabla \Psi = \dot \Psi +i [A,\Psi]$, $\nabla \bar\Psi = \dot {\bar\Psi} +i[A,\bar\Psi]$,
and $\nabla X$, $\nabla Z$ are defined in \p{cov-der}.
It is easy to show that the bosonic limit of \p{S-C-N2c} coincides with the action of the rational Calogero model in the gauge-invariant formulation \p{fed}.
Thus, we have obtained a new ${\cal N}{=}\,2$ supersymmetric extension of the $n$-particle Calogero model with $n$ physical bosons and $2n^2$ fermions $\Psi^b_a$, $\bar\Psi^a_b$, unlike the standard ${\cal N}{=}\,2$ Calogero system with $2n$ fermions proposed in \cite{FrM}.

Note that, after the additional gauge fixing  $Z_a\,{=}\,\bar{Z}^a$, the constrains $(Z_a)^2 \,{=}\,c - R_a$
contain extra fermionic terms $R_a \,{\equiv}\, \{ \Psi , \bar\Psi \}_a{}^a$, $(R_a)^{2n-1}\,{\equiv}\,0$.
At present, it is not clear how to interpret such a proliferation of fermionic fields.
Perhaps, their number could be reduced by implementing a new fermionic gauge invariance similar to the well-known $\kappa$-symmetry.

To deduce ${\cal N}\,{=}\,2$ superextension  of Calogero-Sutherland model, one proceeds from the gauged superfield
sigma-model type action
\begin{equation}\label{S-CS-N2}
S^{N=2}_{CS} = \frac12 \int {\rm d}t {\rm d}^2\theta \,\Big[{\rm tr} \big( {\mathcal X}^{-1}\bar{\mathcal D}
{\mathcal X}
\,{\mathcal X}^{-1} {\mathcal D} {\mathcal X} \big) - \bar {\mathcal Z}\, e^{2V} \,{\mathcal Z} +2c\,{\rm tr} V \,\Big]\,.
\end{equation}
Passing over the same steps as in the rational case, we arrive at the component action
\begin{eqnarray}\label{S-CS-N2c}
S^{N=2}_{CS} & =&  \frac12\int \mathrm{d}t\,\Big[\,{\rm tr}\big(
X^{-1}\nabla X \,X^{-1}\nabla X\big)  + i\, \big(\bar Z \nabla Z - \nabla \bar Z Z\big) + 2c\,{\rm tr} A
\\
&&
+ \, i\,{\rm tr} \big( X^{-1}\bar\Psi X^{-1}\nabla \Psi - X^{-1}\nabla \bar\Psi X^{-1}\Psi \big)
- \, \frac{1}{2}\,{\rm tr} \big( X^{-1}\bar\Psi X^{-1}\bar\Psi X^{-1}\Psi X^{-1}\Psi \big)\Big]\,. \nonumber
\end{eqnarray}
In the bosonic limit, it becomes the gauge-invariant action of the Calogero-Sutherland model \p{L-CS-N0}.
An alternative superspace formulation of both models has been developed in \cite{KLS-19}.

\section{Many-particle ${\cal N}\,{=}\,4$ supersymmetric systems}

The universal approach to superfield formulations of ${\cal N}\,{=}\,4$
mechanics models is the method of ${\cal N}\,{=}\,4$, $d\,{=}\,1$ harmonic superspace \cite{IL},
which is $d\,{=}\,1$ version of the ${\cal N}\,{=}\,2$, $d\,{=}\,4$ harmonic superspace \cite{GIKOS}.
Unlike the ordinary ${\cal N}\,{=}\,4$, $d\,{=}\,1$ superspace with the coordinates $(t, \theta_i, \bar\theta^k)$,
the ${\cal N}\,{=}\,4$, $d\,{=}\,1$ harmonic superspace is parameterized by the coordinates $(t,\theta^\pm, \bar\theta^\pm, u_i^\pm)$, where $\theta^\pm\,{=}\,\theta^i u_i^\pm$,
$\bar\theta^\pm\,{=}\,\bar\theta^i u_i^\pm$, and $u_i^\pm$, $ u^{+i}u_i^-\,{=}\,1$ are ${\rm SU}(2)$-harmonics which parameterize 2-sphere
$S^2 \,{\sim}\, {\rm SU}(2)_R/{\rm U}(1)_R$.
An important property of harmonic superspace is that it has a harmonic analytic subspace, including only half the original Grassmann variables,
$(\zeta,u)\,{=}\,(t_A,\theta^+, \bar\theta^+, u_i^\pm)$, $t_A\,{=}\,t+i(\theta^+\bar\theta^-
+\theta^-\bar\theta^+)$. This analytic superspace is closed under ${\cal N}\,{=}\,4$ supersymmetry.

All ${\cal N}\,{=}\,4$, $d\,{=}\,1$ multiplets can be described by harmonic superfields.
In particular, the ${\cal N}\,{=}\,4$ multiplet $({\bf 1, 4, 3})$ can be represented as a  real harmonic superfield ${\mathcal X}(t,\theta^\pm, \bar\theta^\pm, u)$
subjected to certain constrains (see details in \cite{IL}), or as an analytic prepotential ${\cal V}(\zeta,u)$ defined through the integral representation
\begin{equation}\label{N4-con-sol}
{\mathcal X}(t,\theta_i,\bar\theta^i)=\int du
\,{\mathcal V}(t_A,\theta^+,\bar\theta^+,u)\Big|_{\theta^\pm=\theta^i u^\pm_i,\,\,\,
\bar\theta^\pm=\bar\theta^i u^\pm_i}\,,
\end{equation}
up to gauge transformations $\delta \mathcal{V}\,{=}\,D^{++}\lambda^{--}$,
with the local analytic parameter $\lambda^{--}(\zeta,u)$.
In this section, we also use the ${\cal N}\,{=}\,4$ hypermultiplet described by complex analytical superfields
${\mathcal{Z}}^+ $, $\bar{\mathcal{Z}}^+ $ subjected to the constraint ${D}^{++} \,{\mathcal{Z}}^+\,{=}\,0$,
where $D^{++} \,{=}\,u^{+i}{\partial}/{\partial u^{-i}} +2i\theta^+\bar\theta^+ \partial_{t_A}$ is the analyticity-preserving harmonic derivative (in the analytical basis).
Gauge fields are accommodated  by the unconstrained analytic gauge prepotential $V^{++}$.
Gauge transformation are realized on this superfield as
\begin{equation}\label{N4-gauge-tr}
V^{++}{}^{\,\prime} =  e^{i\lambda}\, V^{++}\, e^{-i\lambda} - i\, e^{i\lambda} (D^{++}
e^{-i\lambda}),
\end{equation}
where $ \lambda_a^b(\zeta, u^\pm)\,{\in}\, u(n) $ is an Hermitian analytic matrix parameter.
Using this gauge freedom, we can choose the Wess-Zumino gauge $V^{++} \,{=}\,2i\,\theta^{+}  \bar\theta^{+}A(t_A)$.

\subsection{${\cal N}\,{=}\,4$ supersymmetric Calogero model}

The matrix superfield action
 \begin{equation}\label{N4-act-0}
S^{N=4}=S^{N=4}_{\mathcal X} + S^{N=4}_{WZ} + S^{N=4}_{FI}
\end{equation}
possesses the most general ${\cal N}\,{=}\,4$, $d\,{=}\,1$ superconformal symmetry $D(2,1;\alpha)$ provided that the items in \p{N4-act-0} are of the form
\begin{equation}\label{N4-act-s}
\begin{array}{l}
{\displaystyle S^{N=4}_{\mathcal X} = {\frac{1}{4(1{+}\alpha)}}\int\! \mu_H  \big({\rm tr} {\mathcal X}^2\big)^{-\frac{1}{2\alpha}},\qquad
S^{N=4}_{WZ}  ={\frac{1}{2}}\int\!
\mu^{(-2)}_A {\mathcal V}_0 \widetilde{\,{\mathcal Z}}{}^{a\,+}
{\mathcal Z}^+_a, }\\ [8pt]
{\displaystyle S^{N=4}_{FI} =-{\frac{ic}{2}}\int\! \mu^{(-2)}_A \,{\rm
tr} \,V^{++} ,}
\end{array}
\end{equation}
where $\mu_H$ and $\mu^{(-2)}_A$ are integration measures in the full and analytic harmonic superspaces.
All superfields in  \p{N4-act-s} are defined by the constraints employing derivatives which are covariant with respect to local $\mathrm{U}(n)$-transformations,
\begin{equation}\label{N4-ga-tr-s}
{\mathcal X}^{\,\prime} =  e^{i\lambda} {\mathcal X} e^{-i\lambda} , \qquad
{\mathcal Z}^+{}^{\prime} = e^{i\lambda} \mathcal{Z}^+\,, \qquad  \bar{{\mathcal Z}}^{+}{}^{\prime} = \bar{\mathcal{Z}}^{+ }e^{-i\lambda} \,,
\end{equation}
e.g.,  $D^{++}{\mathcal Z}^+  \rightarrow {\cal D}^{++}{\mathcal Z}^+ \,{=}\, D^{++}{\mathcal Z}^+ + i V^{++}{\mathcal Z}^+$.
In addition, the superfield ${\cal V}_0$ is a real analytic prepotential for the ${\rm U}(n)$-singlet
superfield ${\mathcal X}_0 \,{\equiv}\, {\rm tr} \left( {\mathcal X} \right)\,$. They are related by the integral transform \p{N4-con-sol}.

Consider the choice $\alpha\,{=}\,{-}1/2$, for which $D(2,1;\alpha)\,{\sim}\, osp(4|2)\,$.
In Wess-Zumino gauge and after eliminating a part of the auxiliary fields, the action \p{N4-act-0} takes the form
\begin{eqnarray}
S^{N=4}_{C}  &=&  \frac12\int dt\big[{\rm tr} \big( \nabla X\nabla X +2c \,A \big)
+ {\textstyle\frac{n}{4}}(\bar Z^{(i} Z^{k)})(\bar Z_{i} Z_{k}) + i\,X_0 \big(\bar Z_k \nabla Z^k
- \nabla \bar Z_k \, Z^k\big)
\big] \nonumber\\
&& + \; \frac{i}{2}\,{\rm tr} \int dt \big( \bar\Psi_k \nabla\Psi^k
-\nabla\bar\Psi_k \Psi^k
\big)  -\int dt  \,\frac{\Psi^{(i}_0\bar\Psi^{k)}_0 (\bar Z_{i}
Z_{k})}{2X_0}\,, \label{S-CS-N4c}
\end{eqnarray}
where $X_0 \,{:=}\, {\rm tr} (X)$, $\Psi_0^i \,{:=}\, {\rm tr} (\Psi^i)$, $\bar\Psi_0^i \,{:=}\,{\rm tr} (\bar\Psi^i)$.
After gauge-fixing  of the residual gauge symmetry, eliminating the fields $A_a^b$, $a \,{\neq}\, b$, and a proper field redefinition,
the bosonic part of the action can be written as
\begin{equation}\label{N4-act-c}
S^{N=4}_{C,b} = \frac12\int dt \Big\{ \sum_{a} \dot x_a \dot x_a + i\sum_{a} (\bar
Z_k^a \dot Z^k_a - \dot {\bar Z}{}_k^a Z^k_a)
-  \sum_{a\neq b} \, \frac{{\rm tr}(S_a S_b)}{4(x_a - x_b)^2} - \frac{n\,{\rm
tr}(\hat S \hat S)}{2(X_0)^2}\,\Big\},
\end{equation}
where $(S_a)_i{}^j \,{:=}\, \bar Z^a_i Z_a^j$, $(\hat S)_i{}^j \,{:=}\, \sum_a \left[ (S_a)_i{}^j -
{\textstyle\frac{1}{2}}\delta_i^j(S_a)_k{}^k\right]$ and
the fields $Z^k_a$ obey the constraints $\bar Z_i^a Z^i_a {=}\,c$ (for any $a$).
The Wess-Zumino term for $Z$-variables in \p{N4-act-c} generates Dirac brackets $[\bar Z^a_i, Z_b^j]_{{}_D}{=}\, i\delta^a_b\delta_i^j$, which as a consequence imply the relation
\begin{equation}\label{N4-alg}
[(S_a)_i{}^j, (S_b)_k{}^l]_{{}_D}= i\delta_{ab}\left\{\delta_i^l(S_a)_k{}^j-
\delta_k^j(S_a)_i{}^l \right\}.
\end{equation}
In other words, for each value of the index $a$, the quantities $S_a$ form mutually commuting algebras $u(2)$, and $(\hat{S})^j_i$
is the conserved Noether ${\rm SU}(2)$-charge of this system.

Unlike the ${\cal N}\,{=}\,2$ cases, not all out of the $d\,{=}\,1$ fields $Z^i_a$
turn out to be auxiliary: after quantization, they become ${\rm U}(2)$-spin degrees of freedom (i.e. harmonics in the target space).
In addition, the quantity $tr \hat S \hat S$ is an integral of motion that generates in the ${\cal N}{=}\,4$ case a conformal potential in the center-of-mass sector.
Modulo this extra conformal potential, the bosonic limit of the  ${\cal N}{=}\,4$ system constructed coincides with the integrable $\mathrm{U}(2)$-spin
Calogero model \cite{Poly06}.

There exists other types of superextensions of the $n$-particle
Calogero model with $su(n)$ or $so(n)$ spin variables
\cite{KLS-18,KLPS-19}. Here, the $su(n)$ spin variables can be
removed by a Hamiltonian reduction, keeping only the ${\cal N}n^2$
fermions for any number $\cal N$ of supersymmetries.

\subsection{${\cal N}\,{=}\,4$ Calogero-Sutherland models}

The main distinguishing feature of this system is the choice of the non-linear sigma-model type action for ${\mathcal X}$ in \p{N4-act-0},
\begin{equation}\label{N4-X-act1}
\tilde S^{N=4}_{\mathcal X} ={\frac{1}{2}}\int \mu_H \,{\rm tr}\, \Big( \ln {\mathcal X}\, \Big),
\end{equation}
with preserving the form of two other terms in
\p{N4-act-0}, \p{N4-act-s}. The full structure of the component action is restored
by the same procedure as in the case of rational Calogero. The
number of physical fermions is again $4n^2$. The action
\p{N4-X-act1} has only ``flat'' ${\cal N}=4$, $d=1$ supersymmetry
and ${\rm SU}(2)$ $R$-symmetry.

Introducing new variables $q_a$ through the replacement $x_a = {\rm e}^{\,q_a}$ brings the bosonic part of the action to the form
\begin{equation}\label{N4-act-com}
S^{N=4}_{CS,b} = \tilde{S}^{N=4}_{CS,b} + \int {\rm d}t \sum_{a,b} \, \frac{(S_a)^{(ik)}\,(S_b)_{(ik)}\,{\rm tr}\big(X^2\big)}{4(X_0)^2} \,,
\end{equation}
where ${\rm Tr}\left(X^2\right)\,{=}\,\sum_c {\rm e}^{2q_c}$, $X_0\,{=}\,\sum_c {\rm e}^{q_c}$,
the constraints $\bar Z_i^a Z^i_a\,{=}\,c$ are satisfied for each $a$ and
\begin{equation}\label{N4-act-comp1}
\tilde{S}^{N=4}_{CS,b} = \frac12\, \int {\rm d}t \Bigg\{ \sum_{a} \,\big[\, \dot q_a \dot q_a +
i (\bar Z_k^a \dot Z^k_a - \dot {\bar Z}{}_k^a Z^k_a) \big]
- \sum_{a\neq b} \, \frac{(S_a)_i{}^k(S_b)_k{}^i}{4\sinh^2 \frac{q_a - q_b}{2}}\Bigg\}\,.
\end{equation}
Therefore, modulo the last term, the action \p{N4-act-com} describes the hyperbolic $\mathrm{U}(2)$-spin Calogero-Sutherland system \cite{Poly06}.

The choice of the action $S_{WZ}$ in \p{N4-act-s} for ${\cal N}\,{=}\,4$ rational Calogero model was mainly motivated by superconformal invariance.
In the hyperbolic case, such symmetry is absent from the very beginning.
In particular, the action \p{N4-X-act1} for ${\cal X}$ has no longer this invariance, and there is no reason to insist on it in other parts of the total action.
Therefore, in the  action \p{N4-act-0} it is natural to choose, instead of \p{N4-act-s}, the simplest action for the multiplets $({\bf 4, 4, 0})$
\begin{equation}\label{N4-Z}
\tilde S^{N=4}_{WZ} = -\frac12 \int \mu_A^{(-2)} \bar{\cal Z}^{+ a}{\cal Z}^{+}_{a}\,.
\end{equation}
The new total action in its bosonic sector yields the ``pure''
hyperbolic $\mathrm{U}(2)$-spin Calogero-Sutherland system for any
$n$, without any additional interaction. The coordinate of the
center of mass is completely separated and is described by a free
action in this model.

Also for the Calogero-Sutherland models, there exists an ${\cal
N}{=}\,4$ supersymmetric extension free of spin variables
 and still containing $4n^2$ fermions \cite{KL-20}.

\section{Conclusions}

We have described a universal method of constructing supersymmetric extensions of Calogero-type models based on the superfield gauging procedure.
This method leads to a non-standard supersymmetrization with ${\cal N}n^2$ physical fermionic fields.
Using it, we constructed new ${\cal N}\,{=}\,2$ and ${\cal N}\,{=}\,4$ superfield systems containing rational Calogero models
and hyperbolic Calogero-Sutherland models as a bosonic limit for ${\cal N}\,{=}\,2$ case and their $\mathrm{U}(2)$-spin analogs for
${\cal N}\,{=}\,4$ case.

We finish by listing some further possible tasks in the framework of the approach proposed: \\

\noindent $\bullet$\;Studying the classical and quantum integrability of new supersymmetric Calogero models; \\

\noindent $\bullet$\; Considering the possibilities of representing spin variables in various ${\cal N}\,{=}\,4$ Calogero systems
by other ${\cal N}=4$, $d=1$ multiplets, for example, multiplets (${\bf 2, 4, 2})$ or $({\bf 3,4,1})$;  \\

\noindent $\bullet$\;Generalization of the gauge approach to the case of ${\cal N}\,{=}\,4$ ``weak'' supersymmetries $\mathrm{SU}(2|1)$ \cite{BN03,Sm04,IvS14}
and similar deformed versions of ${\cal N}\,{=}\,8$ supersymmetry \cite{ILS16}, with some additional oscillator-type terms; \\

\noindent $\bullet$\;Quantization of all these models like it was recently done in \cite{FILS18} for Calogero-Moser systems with $\mathrm{SU}(2|1)$ supersymmetry; \\

\noindent $\bullet$\;Reproducing, by the superfield gauging method, the multiparticle systems constructed in \cite{KLS-18,KLPS-19} in the Hamiltonian on-shell approach for arbitrary ${\cal N}$; \\

\noindent $\bullet$\;Supersymmetrizing other integrable many-particle models from the list of \cite{OP1}, e.g., trigonometric Calogero-Sutherland models, elliptic models, etc.

\vspace{0.5cm}
\noindent {\large\bf Acknowledgments} \\

\noindent E.I. and S.F. thank the Russian Science Foundation, grant No.\,16-12-10306, for a partial support.

\end{document}